\newcommand{\manuscript}{0}

\ifnum\manuscript=2
\documentclass[11pt, letterpaper]{article}
\newcommand{\shorttitle}[1]{}
\newcommand{\shortauthors}[1]{}
\newcommand{\received}[1]{}

\newcommand{\acknowledgements}[1]{ }
\renewcommand{\author}[1]{#1}
\renewcommand{\title}[1]{{\centering {\Large \bf #1}\\ }}

\usepackage[margin=1.2in]{geometry}
\else\ifnum\manuscript=1
\documentclass[12pt,letterpaper,preprint,flushrt]{aastex}
\else
\documentclass[prl,twocolumn]{revtex4}
\newcommand{\comment}[1]{}
\fi\fi

\usepackage{amsmath,aas_macros}

\newcommand{\commentx}[1]{}

\renewcommand{\vec}[1]{\mbox{\boldmath$#1$}} 

\newcommand{\ra}[3]   
   {\makebox[1.5em][r]{#1}\makebox[1.5em][r]{#2} \makebox[2em][r]{#3}}

\newcommand{\hms}[3]  
   {${#1}^{\mathrm{h}}{#2}^{\mathrm{m}}{#3}^{\mathrm{s}}$}

\newcommand{\hmin}[2]  
   {\ensuremath{{#1}^{\mathrm{h}}{#2}^{\mathrm{m}}}}
   
\newcommand{\hours}[1]  
   {\ensuremath{{#1}^{\mathrm{h}}}}

\newcommand{\dms}[3]  
   {\ensuremath{{#1}\degree{#2}\arcminute{#3}\arcsecond}}

\newcommand{\dm}[2]  
   {\ensuremath{{#1}\degree{#2}\arcminute}}

\newcommand{\ukcmb}  
           {\ensuremath{\micro \kelvin_\mathrm{cmb}}}

\newcommand{\uk}  
           {\ensuremath{\micro \kelvin}}

\newcommand{\fdeg} 
           {\hbox{$.\!\!^{\circ}$}}

\usepackage{color}

\newcommand\beq{\begin{equation}}
\newcommand\eeq{\end{equation}}
\newcommand\beqn{\begin{eqnarray}}
\newcommand\eeqn{\end{eqnarray}}
\newcommand\ave[1]{\left\langle {#1} \right\rangle}

\newcommand{\bm}[1]{\ensuremath{\mbox{\boldmath $#1$}}} 
\newcommand\bl{{\bm{\ell}}}

\newcommand\bL{{\mathbf{L}}}

\newcommand\nn{\nonumber}

\newcommand{\cmbav}[1]{\ave{#1}_{\mathrm{CMB}}}
\newcommand{\lssav}[1]{\ave{#1}_{\mathrm{LSS}}}
\newcommand{\modu}[1]{\left |{#1}\right |}

\usepackage{graphicx,hyperref,SIunits}

\begin{document}
\title{CMB Lensing --- Power Without Bias}
\author{Blake D.~Sherwin}
\email{bsherwin@princeton.edu}
\affiliation{Department of Physics, Jadwin Hall,  Princeton University, Princeton, NJ 08544}
\author{Sudeep Das}
\email{sudeep.das@berkeley.edu}
\affiliation{Berkeley Center for Cosmological Physics,  Department of Physics, University of California, Berekeley, CA 94720}
\begin{abstract}

We propose a novel bias-free method for reconstructing the power spectrum of the weak lensing deflection field from cosmic microwave background (CMB) observations.  The proposed method is in contrast to the standard method of CMB lensing reconstruction where a reconstruction bias needs to be subtracted to estimate the lensing power spectrum. This bias  depends very sensitively  on the modeling of the signal and noise properties of the survey, and a misestimate can lead to significantly inaccurate results.   Our method obviates this bias and hence the need to characterize the detailed noise properties of the CMB experiment. We illustrate our method 
with simulated lensed  CMB maps with realistic noise distributions. This bias-free method can also be extended to create much more reliable estimators for other four-point functions in cosmology, such as those appearing in  primordial non-Gaussianity estimators. 
\end{abstract}
\maketitle

When the universe was $\sim$380,000 years old, the photons of the cosmic microwave background (CMB) decoupled from the primordial photon-baryon fluid. Traveling 
through the newly transparent universe, the photons were deflected many times by the gravitational influence of the large-scale structure potentials through
which they passed, an effect  known as the gravitational lensing of the CMB, or CMB lensing (for a review, see \cite{lewis/challinor:2006}). The net effect of lensing is an effective deflection of each CMB photon, leading to a remapping of the points on the CMB sky. The deflection field  depends  on a distance-weighted projection of  density perturbations along the line of sight. The power spectrum of  the deflection field is  therefore sensitive to both geometry and the growth of structure over a broad redshift range ($z\sim 0.5 -5 $). As such, 
knowledge of the convergence field can provide strong constraints
on parameters that affect geometry or the growth at later times, such as the sum of neutrino masses and parametrizations of non-standard dark energy behavior \citep{smith/etal:prep}.  These constraints are complementary to those obtained directly from the primordial CMB anisotropies. In the very near  future, ongoing and upcoming CMB  experiments, such as Planck, ACT, SPT, PolarBear, ACTPol and SPTPol will produce datasets with sufficient resolution and sensitivity to 
begin the determination of the deflection field and ultimately realize the cosmological potential of CMB lensing science. Robust algorithms that are insensitive to the details of the noise properties of the survey will be essential for accurate determination of the deflection power spectrum.  \par
In this Letter, we propose a novel and \emph{bias-free} technique for the measurement of the lensing
deflection  power spectrum. This is in contrast to 
the standard optimal quadratic estimator (OQE) method where a bias term, comparable to and often much larger than the  signal, has to be subtracted (for details, see \cite{hu/okamoto:2002, kesden/cooray/kamionkowski:2003}).  This bias, which is a temperature-field four point function that depends on noise and foregrounds, must typically be computed or simulated
to an accuracy of a few percent in order to get reliable estimation of the signal. Because of our limited knowledge of the temperature and polarization foregrounds, and the
typically complicated noise properties of CMB experiments, modeling this bias 
term sufficiently accurately for a robust detection of lensing may not be possible. 
\par
To discuss how this bias appears and can be avoided, we will review some lensing theory
here.  The lensed and unlensed temperature fields (unlensed quantities will be denoted by a tilde) are related by $T(\mathbf{\hat{n}}) = \tilde{T}(\mathbf{\hat{n}+\vec{\alpha}(\hat{n})})$, where $\alpha(\mathbf{\hat{n}}) $ is the deflection field and $\kappa = \frac{1}{2} \nabla \cdot \vec \alpha$ is the convergence field. In the flat sky approximation, the temperature field can be expanded as a Taylor series in Fourier space \citep{zaldarriaga/seljak:1998, seljak/zaldarriaga:1999}:
\beq
\label{kappa}
T(\bl) = \tilde{T}(\bl)+2 \int \frac{d^2\bl'}{(2\pi)^2} \frac{\bl'\cdot (\bl-\bl')}{|\bl-\bl'|^2} \kappa(\bl-\bl')\tilde{T}(\bl') + O(\kappa^2).
\eeq
Thus, gravitational lensing introduces correlations between the formerly independent modes of the temperature field, which can be used to construct a quadratic estimator for $\kappa$:
\begin{eqnarray}
\label{kappaEst}
\nonumber \hat\kappa(\bL) &=& N^{\kappa}(\bL)   \int  \frac{d^2\bl}{(2\pi)^2}  ~ \bL \cdot  (\bL - \bl)  \\ &&\times   F_{W}(\bl) {T(\bl)}{   F_{G}(\modu{\bL-\bl}) }T(\bL-\bl). 
\end{eqnarray}
where $N^{\kappa}$ is a normalization that ensures  that the estimator is unbiased (i.e., $\cmbav{\hat\kappa} =\kappa$) and $F_W$ and $F_G$ are filters that can be tuned to 
minimize its variance \cite{hu/okamoto:2002}.  Here we will use notation from \cite{kesden/cooray/kamionkowski:2003}  and denote ensemble averages over CMB realizations with the large scale structure (LSS) fixed as $\cmbav{...}$, and averages over LSS realizations as $\lssav{...}$, and  use $\ave{...}$ to denote  ensemble averages over both, i.e., $\lssav{\cmbav{...}}$.

While equation (\ref{kappaEst}) provides an unbiased estimate of the convergence field, the naive lensing power spectrum estimator $\hat\kappa^*(\bL)\hat\kappa(\bL)$ is highly biased:
\beqn
\nn\ave {\hat\kappa^*(\bL)\hat\kappa(\bL')} &=&   N^{\kappa *}(\bL) N^{\kappa }(\bL')\\
\nn &&  \int  \frac{d^2\bl}{(2\pi)^2}   \int  \frac{d^2\bl'}{(2\pi)^2}g(\bl,\bL) g(\bl',\bL')  \\
\label{twoDInt}   && \times \scriptsize \lssav{ \cmbav{T^*(\bl)~T^*(\bL-\bl)~ T(\bl')   ~T(\bL'-\bl')}}\\
\nn &\simeq &  (2 \pi)^2 \delta(\bL-\bL')  C_L^\kappa \\
\nn && +   ~ N^{\kappa *}(\bL) N^{\kappa }(\bL')\\
\nn && \times \int  \frac{d^2\bl}{(2\pi)^2}   \int  \frac{d^2\bl'}{(2\pi)^2}g(\bl,\bL) g(\bl',\bL')  \\
  && \times \scriptsize\cmbav{\tilde T^*(\bl)~\tilde T^*(\bL-\bl)~ \tilde T(\bl')  ~\tilde T(\bL'-\bl')}
\eeqn
where $g(\bl,\bL) = \bL \cdot  (\bL - \bl) F_{W}(\bl) {   F_{G}(\modu{\bL-\bl}) } $. In going from the first equality above to 
the second, we have neglected a few terms that appear involving integrals over the convergence power spectrum. 
These higher order terms are computed in   \cite{kesden/cooray/kamionkowski:2003} and are subdominant 
compared to the Gaussian four point term above. Applying Wick's theorem contractions to the four-point term, the above equation can be reduced to:
\beq
\nn\ave {\hat\kappa^*(\bL)\hat\kappa(\bL')}   = (2 \pi)^2 \delta(\bL-\bL') ( C_L^\kappa +C_L^{\kappa,\rm Gauss}),
\eeq
where 
\beqn
\label{gaussBias}
\nn C_{\bL}^{\kappa,\rm Gauss} &=& N^{\kappa*}(\bL) N^\kappa(\bL)  \int \frac{d^2\bl}{(2\pi)^2} \int \frac{d^2\bl'}{(2\pi)^2} f(\bl,\bl',\bL) \\
\nn && \left[ C_{\bl} C_{\bL-\bl} (2 \pi)^2 \delta(\bl'-\bl)\right.\\&&
 \left.+C_{\bl} C_{\bL-\bl} (2 \pi)^2 \delta  \left(\bl'-(\bL-\bl) \right) \right],
\eeqn
with $ f(\bl,\bl',\bL) =g(\bl,\bL)~g(\bl',\bL')$. This expression shows that the Gaussian bias $C_{\bL}^{\kappa,\rm Gauss}$ depends on the map power 
spectrum estimate which is usually a sum of the CMB temperature power spectrum, foregrounds, and  noise.  Since the Gaussian bias term is typically more than an order of magnitude larger than the intrinsic lensing signal $C_\ell^{\kappa}$, the standard approach to estimating lensing requires a detailed understanding of each of these contributions. \par
\begin{figure}[t]
\label{bd}
 \centering
 \includegraphics[width=3.4in,bb=0 0 750 586]{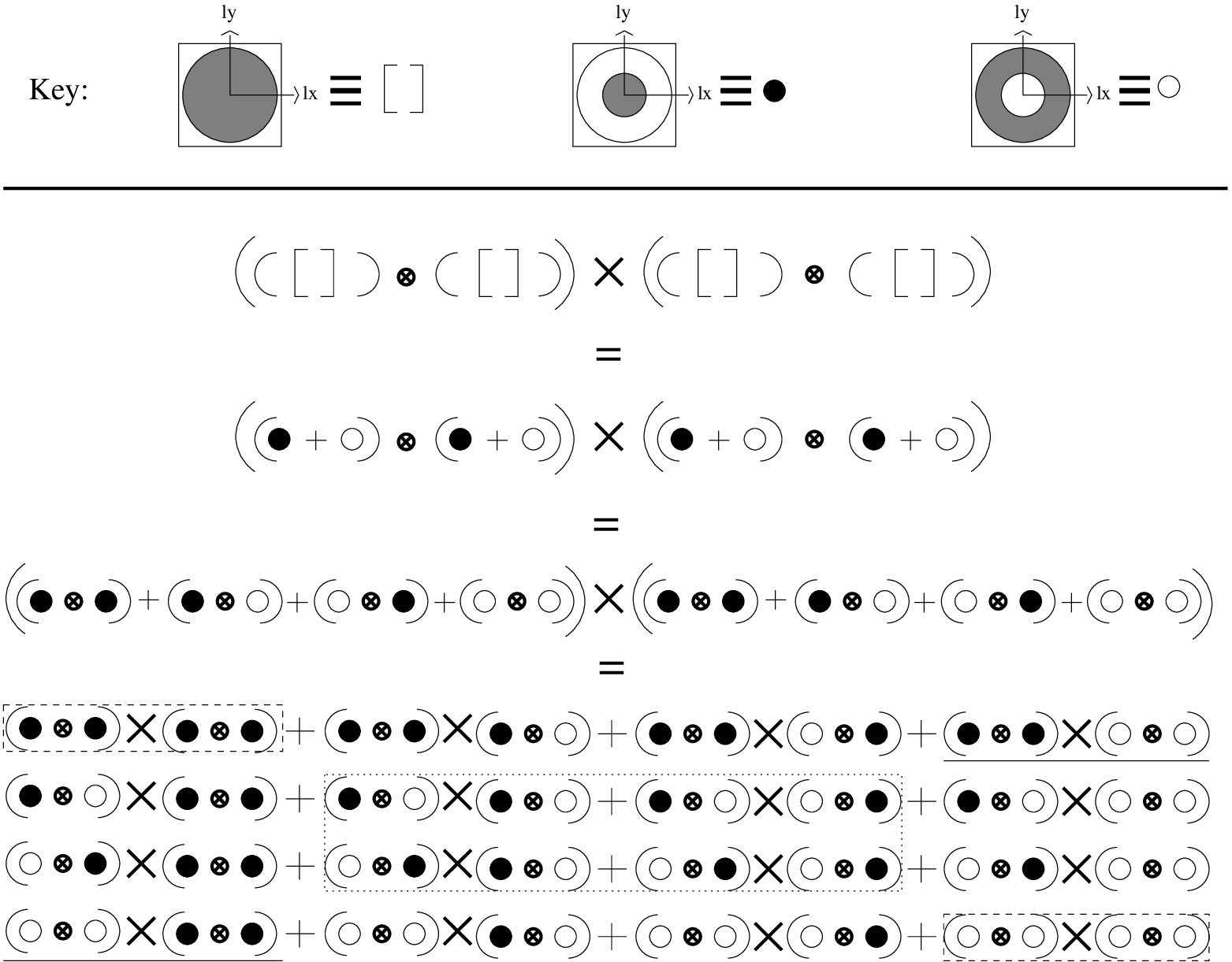}
\caption{Graphical expansion of the naive estimator after splitting up the Fourier space into an inner and an outer annulus. We use the linearity of both operators in this expansion. The terms with Gaussian bias are shown enclosed by boxes. The  underlined terms  (identical by symmetry) are implemented in the simulations described  in this paper to illustrate the method.   \label{schematic}}
\end{figure}
The goal of this paper is to eliminate the need to compute this Gaussian bias term at the percent level by eliminating the bias altogether. From (\ref{twoDInt}) and (\ref{gaussBias}), one can see that the brute force way to achieve this would be to perform the double two-dimensional integral explicitly in  (\ref{twoDInt})  with the following conditions imposed on the function $f$:
\beq
f(\bl,\bl',\bL) = 
\begin{cases} 0 & \text{if $\bl'=\bl$ or $\bl' = \bL-\bl$}
\\
f(\bl,\bl',\bL)  &\text{otherwise.}
\end{cases}
\eeq
However, this method is computationally expensive and not efficient.   A conceptually simpler  and
much more efficient method of eliminating the bias would be  to partition the Fourier space into non-overlapping annuli, and cross correlate the $\kappa(\bL)$'s  reconstructed from temperature modes in two disjoint annuli.  
\begin{figure*}[!ht]
\label{bd}
 \centering
 \includegraphics[scale=0.45]{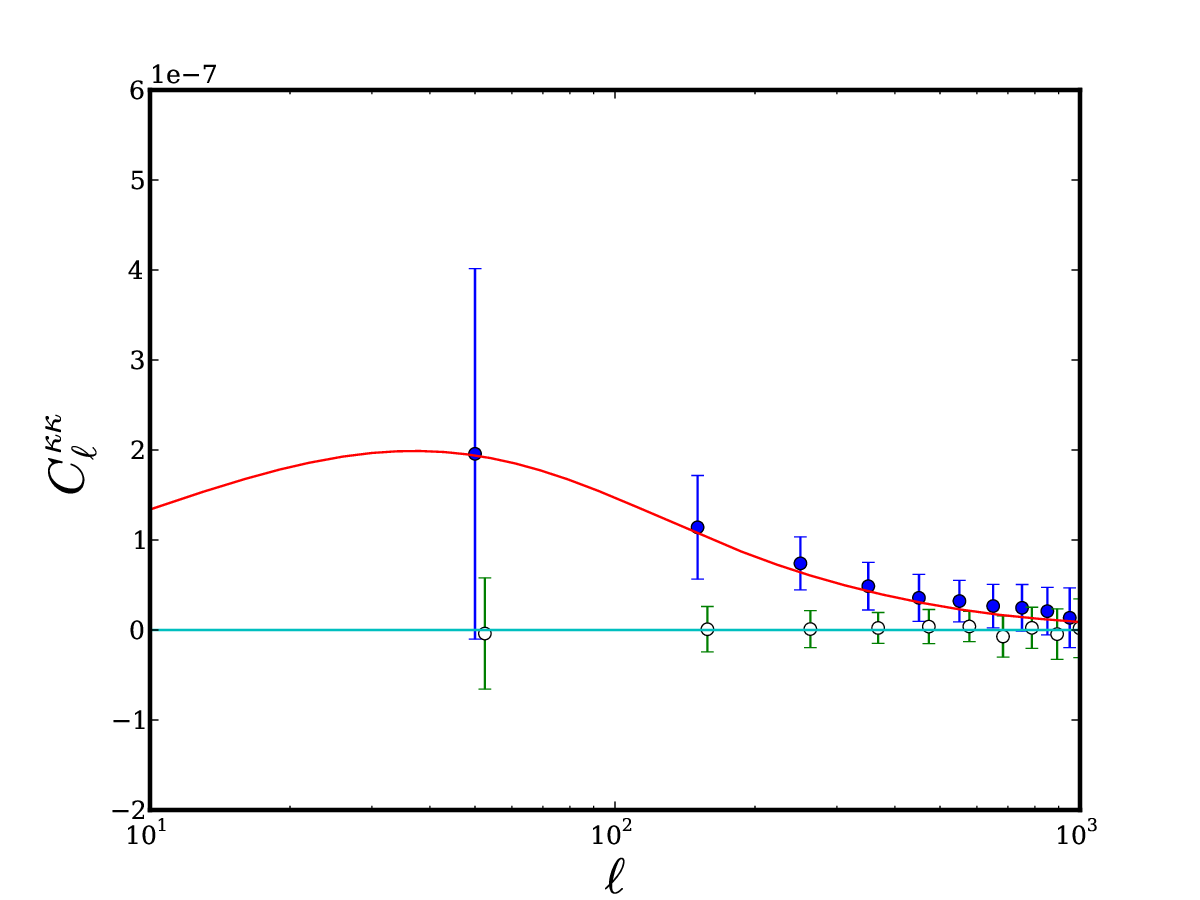}\includegraphics[scale=0.45]{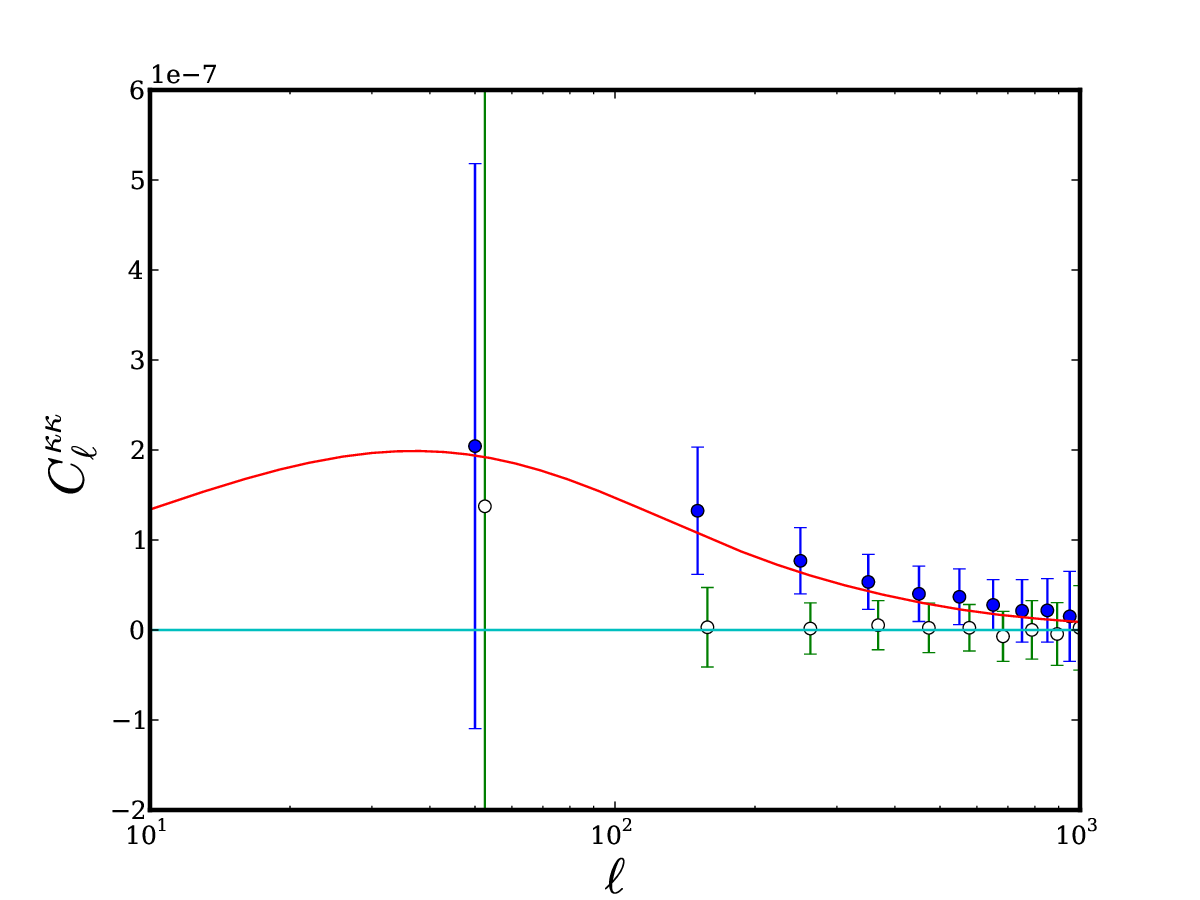}
\caption{\emph{Left:}  Convergence power spectrum reconstructed with the proposed Gaussian bias-free method from four $5\degree\times 15 \degree$  patches with simulated  CMB signal  with 2 $\micro\kelvin$-arcmin white noise. The blue (filled) circles show the mean of 120 Monte Carlo realizations with lensed CMB,  while the green (empty) circles show the same for unlensed CMB maps.   The error bars are estimated from the scatter between Monte Carlo runs and are representative of  the uncertainty expected in one  realization; the lensed errors are higher than the null errors due to the presence of a sample variance component. The red continuous curve is the input theory for 
the convergence field power spectrum.  \emph{Right:} Same as left, but for non-white and anisotropic noise simulations  seeded  by noise in ACT maps, reduced in amplitude by a factor of 3.  \label{simResults}}
\end{figure*}
To formulate this fully, let us
introduce some compact notation.  Let us denote the operation of reconstructing $\kappa$ from two temperature
maps, equation (\ref{kappaEst}), as $\kappa = (T\otimes T)$ where all quantities are understood to be  in Fourier $(\bl)$ space.   Now, the naive estimator for the convergence power spectrum (\ref{twoDInt}) can be written as $ \hat \kappa^* \hat\kappa  =  (T \otimes T) \times (T \otimes T)$. Now consider breaking up $T(\bl)$ into two Fourier
space maps, one which has non-zero elements only with an annulus $\ell_0<\ell<\ell_1$ (the ``in-annulus'') and another which has non-zero values in $\ell_1<\ell<\ell_2$ (the ``out-annulus'' ), where $\ell_0<\ell_1<\ell_2$.  Writing $T(\bl) = T_{\rm in}(\bl) + T_{\rm out}(\bl)$,  the naive power spectrum estimator can be written out as, 
\beqn
\nn \hat C_{\bl}^{\kappa,\mathrm{naive}} &=& ( T \otimes  T) \times ( T \otimes  T) \\
\nn && = (( T_{in}+ T_{out}) \otimes ( T_{in}+ T_{out})) \\
&& \times (( T_{in}+ T_{out}) \otimes ( T_{in}+ T_{out})).
\eeqn
We have expanded this out graphically in Fig.~\ref{schematic}. In this figure, we represent the in-annulus by a filled circle and the out-annulus  by an empty circle. The expansion leads to 16 terms, some of which are identical due to symmetry.  Note that the Gaussian bias associated with each term evaluates to a sum of two terms from two possible Wick's theorem pairings (one member of the pair being taken from either side of the ``$\times$" sign).  Those terms (e.g. $(T_{in} \otimes T_{in}) \times ( T_{out} \otimes T_{out})$, underlined in Fig.~\ref{schematic} ) where either pairing leads to at least one  product of an out-annulus with an in-annulus will have contributions to the bias which evaluate out to zero. We find 10 such terms in Fig.~\ref{schematic}; the remaining 6 terms' biases have non-zero expectation value and hence contribute to the total bias (these terms are shown enclosed by  boxes).  Therefore, one can construct a new estimator for the convergence power spectrum by applying these annular filters and optimally combining the 10 terms  (some of these are identical) that are  bias-free. 
Of course,  eliminating bias comes at the cost of reducing the signal-to-noise because we throw out a fraction of  terms with information. In principle,  higher signal-to-noise can be achieved by further subdividing each annulus into an inner and an outer part,  and iterating the method, thereby reducing the ratio of the number of biased to bias-free terms.  \par

There are a few details that need to be taken into account when applying this method to a real experiment. For a partial sky map, nearby Fourier modes will be coupled by the power spectrum of the data-window with 
some characteristic width $\Delta \ell$. There can be additional effects such as coupling of nearby modes induced by anisotropic noise.  In general, if the effective width $\Delta \ell$ of such correlations is known,  the above method needs to modified by separating the two annuli  by some  small multiple 
of $\Delta\ell$.  This will ensure that our annular method for eliminating the Gaussian bias works despite these correllations due to systematics. 

Due to the annuli being separated by $\Delta\ell$, all the terms involving a convergence map obtained from an innner annulus as well as an outer annulus are undefined for $\ell<\Delta \ell$ (as can be deduced from equation \ref{kappaEst}). Hence, for a simple bias-free estimator we used only the terms underlined in Fig.~\ref{schematic}: those with one ``in--in" convergence map crossed with one ``out--out" convergence map, so that the new bias free estimator is:
\beq
\hat C_{\bl}^{\kappa,\mathrm{bias-free}} = (T_{in} \otimes T_{in}) \times (T_{out} \otimes T_{out})
\eeq

There is one further subtlety to the reconstruction of the convergence power spectrum. Window function correlations due to finite maps and anisotropic noise not only affect the Gaussian bias, but also appear more directly as a spurious lensing signal in the reconstructed convergence maps. This spurious convergence, $\ave{\tilde{\kappa}}$, needs to be simulated and subtracted off from the reconstructed $\hat\kappa$ map. To determine whether the reconstructed convergence power is sensitive to the accuracy of the simulation of $\ave{\tilde{\kappa}}$, we must determine the magnitude of $C_\bl^{\ave{\tilde{\kappa}}_{in,in} \ave{\tilde{\kappa}}_{out,out}}$ (where one must distinguish between the two $\ave{\tilde{\kappa}}$ fields because they are simulated with different annular filters). From our Monte Carlo simulations, which we describe below, we find the following: for our bias-free lensing estimators, $C_\bl^{\ave{\tilde{\kappa}}_{in,in} \ave{\tilde{\kappa}}_{out,out}}$ appears consistent with null and is typically two orders of magnitude smaller than the true convergence power $C_\ell^{\kappa}$ and three orders of magnitude smaller than the previously discussed reconstruction bias. We also verified that the results of our simulations plotted below are the same whether or not we subtract $\ave{\tilde{\kappa}}$ from $\hat \kappa$. The accuracy of the simulation of the spurious signal $\ave{\tilde{\kappa}}$ thus seems to only have a negligible influence on how well the lensing power spectrum is reconstructed.\par
We illustrate our method for bias-free lensing power reconstruction by performing Monte Carlo simulations, loosely modeled on the observations made by the Atacama Cosmology Telescope (ACT). We choose our survey geometry to be an oblong $5 \degree \times 60 \degree$ stripe, divided into four adjacent patches of $5 \degree \times 15 \degree$ each.  We simulate convergence maps on these patches from an input power spectrum, and generate  deflection fields from the  
convergence maps. We also generate Gaussian random realizations of unlensed CMB maps from an input power spectrum, which we subsequently lens using the simulated deflection field.  Then we smooth the maps with 
the ACT beam (1.4 arcmin full-width-half-maximum), and add noise. We perform two variations on the noise. 
  In the first version, we add white noise at the  level of $2 ~\micro\kelvin$-arcmin.  In the second version,
  we simulate noise seeded  by the noise power spectrum realized in the maps  from ACT. These simulations 
  capture the non-white and anisotropic aspects of the noise in ACT  maps (for the detailed procedure   see \cite{das/etal:prep}). To ensure that the reconstructed convergence power spectrum  is not too noisy (so that the Monte Carlo simulations rapidly converge), we reduce the amplitude of the simulated noise by a factor of three  over what is observed in  ACT maps.  These maps are roughly $10 ~\micro\kelvin$-arcmin in noise. For each type of noise, we simulate 120 realizations of the full map by randomizing the CMB, the convergence and the noise. 
  We then apply the bias-free 
  convergence power spectrum estimator method to each random realization of the noisy maps. We apply this
  both to lensed maps with noise and as a null test also to unlensed maps with noise.   We define our annular filters such that the  inner annulus is $\ell=(500-1500)$, the outer annulus is $\ell=(1900-3300)$, so that $\Delta\ell=400$. 

The results are shown in Fig.~\ref{simResults}. This figure shows that in both cases our method is able to 
extract the convergence power spectrum without bias (note that a small amount of bias from higher order terms discussed in \citet{kesden/cooray/kamionkowski:2003} is present in the reconstruction). The figure also shows that
with noisy but unlensed CMB maps, the measurements are  consistent with  a null signal. 
\par 
Finally, we should point out some caveats.  Here we have assumed CMB as the only signal. In reality, emission from point sources  and the thermal and kinetic Sunyaev-Zeldovich effects contribute to mm-wave maps.  Also, noise correlations in real experiments can be  more complicated than what is simulated here. There are also other subdominant sources
of reconstruction bias, such as those discussed in  \cite{kesden/cooray/kamionkowski:2003} and \cite{hanson/etal:prep}.  More work will be needed to characterize these foregrounds and biases in context of our new method. 
These will be discussed in a future, more detailed work. \par
It should be noted that our approach  for bias-free  lensing convergence reconstruction can be easily extended to estimating other four-point-functions in cosmology, such as the estimators of primordial non-Gaussianity.

\acknowledgements
BDS and SD would like to acknowledge fruitful discussions with Kavilan Moodley and thank David Spergel, Kendrick Smith, Lyman Page and Suzanne Staggs for their advice and feedback. BDS also acknowledges the hospitality of the BCCP during the development of this paper.  Computations were performed on the GPC supercomputer at the SciNet HPC Consortium. SciNet is funded by: the Canada Foundation for Innovation under the auspices of Compute Canada; the Government of Ontario; Ontario Research Fund - Research Excellence; and the University of Toronto. This project was partially supported by NSF grant 0707731 and NASA grant NNX08AH30G. BDS is supported by a National Science Foundation Graduate Research Fellowship. SD is supported by a Berkeley Center for Cosmological Physics postdoctoral  fellowship.

\end{document}